\documentclass[twocolumn]{aastex631}
\usepackage{amsmath}
\usepackage{wrapfig}
\usepackage{svg}

\newcommand{\MBH}{M_{BH}}
\newcommand{\Mmw}{M_{\rm MW}}

\newcommand{\trlx}{\tau_{\rm rlx}}

\newcommand{\dd}{{\rm d}}
\newcommand{\rstar}{R_{\star}}

\newcommand{\Msun}{{M_{\odot}}}
\newcommand{\Rsun}{{R_{\odot}}}
\newcommand{\mb}{m_\bullet}
\newcommand{\tmb}{\widetilde{m}_\bullet}
\newcommand{\Nb}{N_\bullet}
\newcommand{\Ns}{N_\star}
\newcommand{\gb}{\gamma_\bullet}

\newcommand{\tgb}{\widetilde{\gamma}_\bullet}
\newcommand{\Rb}{R_\bullet}
\newcommand{\fb}{\xi_\bullet}

\newcommand{\Rate}[1]{\left.\frac{\dd\Gamma}{\dd m}\right|_{#1}}

\usepackage{comment}

\graphicspath{{graphs/}}

\begin{document}

\title{Mass Segregation and Transient Formation in Nuclear Stellar Clusters}

\correspondingauthor{Barak Rom}
\email{barak.rom@mail.huji.ac.il}

\author[0000-0002-7420-3578]{Barak Rom}
\affiliation{Racah Institute of Physics, The Hebrew University of Jerusalem, 9190401, Israel}

\author[0000-0002-1084-3656]{Re'em Sari}
\affiliation{Racah Institute of Physics, The Hebrew University of Jerusalem, 9190401, Israel}

\begin{abstract}
\noindent
Supermassive black holes at the centers of galaxies occasionally disrupt stars or consume stellar-mass black holes (BHs) that wander too close, producing observable electromagnetic or gravitational wave signals. 
We examine how mass segregation impacts the rates and distributions of such events. 
Assuming a relaxed stellar cluster, composed of stars and stellar-mass BHs, we show that the tidal disruption rate of massive stars ($m\gtrsim\Msun$) is enhanced relative to their abundance in the stellar population. 
For stars up to $m\approx3\Msun$, this enhancement is roughly $m/M_\odot$ and it is driven by segregation within the sphere of influence. Stars with masses $m\gtrsim3\Msun$, if relaxed, are predominantly scattered by more massive stellar-mass BHs, leading to a constant enhancement factor of $\approx 9$, independent of mass.
This aligns with observational evidence suggesting an over-representation of massive stars in tidal disruption events.
For stellar-mass BHs, we predict an enhancement factor scaling as $m_\bullet^{1/2}$ for plunges and $\mb^{3/2}$ for extreme-mass-ratio inspirals (EMRIs). The power of one-half in both cases reflects the shorter relaxation times of heavier BHs, allowing them to segregate into the sphere of influence from greater distances, thereby increasing their abundance. 
The additional power in the EMRIs' rate arises from the tendency of heavier BHs to circularize and sink inward more efficiently.
Finally, we estimate the rate of main-sequence star inspirals and find that it favors low-mass stars ($m\lesssim\Msun$). 
This seems compatible with the observationally estimated rate of quasiperiodic eruptions.
\end{abstract}


\section{Introduction}
The centers of galaxies harbor supermassive black holes (SMBHs) and their surrounding nuclear stellar clusters. 
These dense environments give rise to various electromagnetic and gravitational wave (GW) transients, such as tidal disruption events \citep[TDEs;][]{Rees_88,Gezari_21}, quasiperiodic eruptions \citep[QPEs;][]{Miniutti_2019,Arcodia_2021}, merging stellar-mass black hole (BH) binaries \citep{Mapelli_2021,sedda_2023}, and extreme-mass-ratio inspirals \citep[EMRIs;][]{Amaro_2018}.

A star is tidally disrupted if it passes too close to the SMBH, namely closer than its tidal radius, where the SMBH's tidal force and the star's self-gravity are comparable \citep{Hills_1975}
\begin{equation}\label{eq:Rt}
    \begin{aligned}
        R_t&\approx\rstar\left(\frac{\MBH}{m}\right)^{1/3}\\
        &\approx10R_s \left(\frac{m}{\Msun}\right)^{0.47} \left(\frac{\MBH}{\Mmw}\right)^{-2/3},
    \end{aligned}
\end{equation}
where $m$ and $\rstar$ are the mass and the radius of the star, respectively, and $R_s=2G\MBH/c^2$ is the SMBH Schwarzschild radius.
We normalize $\MBH$ to the mass of Sgr A*, $\Mmw= 4\times10^6$ \citep{Ghez_2008,Gillessen_2009},  and use the main-sequence mass-radius relation, $\rstar\approx\Rsun (m/
\Msun)^{0.8}$. 

Observationally, about $100$ TDEs have been discovered \citep{Gezari_21,Hammerstein_2023,Yao_2023}, primarily in the past decade, through wide-field surveys, such as the All-Sky Automated Survey for Supernovae \citep{Jayasinghe_2018} and the Zwicky Transient Facility \citep{Bellm_2019}. 
The number of detected TDEs will increase significantly with the upcoming Vera Rubin Observatory Legacy Survey of Space and Time \citep{Ivezic_2019}, which is expected to detect $\sim10^3$ TDEs per year \citep{van_Velzen_2011,Bricman_2020}.

Furthermore, the recent discoveries of QPEs \citep{Miniutti_2019,Arcodia_2021} uncovered a new class of transients from centers of galaxies, characterized by repeating X-ray emissions over periods of hours. The growing number of observed QPEs reveals the complexity and diversity of phenomena associated with these events \citep{Arcodia_2022,Miniutti_2023a,Miniutti_2023b,Arcodia_2024b,Chakraborty_2024}. 
Several formation models have been proposed, including mass transfer from white dwarfs \citep[WDs;][]{King_22} or stars \citep{MetStoGil22,LuQua_2023,Linial_2023}, star-disk interactions \citep{Xian_2021,Franchini_2023,Linial_Metzger_2023}, and disk instabilities \citep{Raj_2021,Pan_2022,Kaur_2023}.
Evidence supporting the TDE--QPE connection, as suggested by \cite{Linial_Metzger_2023}, has been observed \citep{Chakraborty_2021,Quintin_2023,Bykov_2024}. Most recently, \cite{Nicholl_2024} provided a direct link between these phenomena by detecting QPEs following a known TDE.

In parallel, BHs accumulate in galactic centers and migrate inward due to dynamical friction, eventually merging with the central SMBH. These mergers produce GW signals in the mHz band, detectable by next-generation, space-based GW observatories, such as LISA \citep{LISA_2017,Amaro_2023} and TianQin \citep{Luo_2016}. 
For these mergers, the minimal distance from the SMBH is $4R_s$, corresponding to the angular momentum of the mostly bound orbit \citep{Hopman_2005}, rather than the tidal radius associated with the disruption of stars. 
BHs that reach a smaller periapsis will rapidly plunge into the SMBH. In contrast, BHs on highly eccentric orbits with larger periapsis distance can dissipate energy via GW emission, circularize, and slowly descend toward the SMBH - such orbital evolution is known as EMRI \citep{Hopman_2005,Amaro_2018}.

The rates and characteristics of these transients depend on the distribution of stars and BHs around the SMBH, which has been extensively studied over the past half a century, since the pioneering works of \cite{Peebles_72} and \cite{BW_76}, both theoretically \citep[e.g.,][]{AH_09,Linial_2022} and numerically \citep[e.g.,][]{Panamarev_2019,Zhang_2025}, as further described in Section \ref{sec:2}.

The TDE rate, $10^{-4}-10^{-5}\ \rm yr^{-1}$ \citep{Magorrian_1999,Wang_2004,Holoien_2015,Stone_2016,Kochanek_2016,van_Velzen_2018,Stone_2020,Bortolas_2024,Yao_2023}, is determined by the replenishment of stars into highly eccentric orbits with periapsis $r_p\lesssim R_t$, mainly via two-body scatterings. 
However, other mechanisms can alter the TDE rate; for example, strong scatterings may reduce the TDE rate, given a steep profile density \citep{Teboul_2024,KaurPerets_2024}, while the presence of an SMBH binary can temporarily increase the TDE rate\footnote{As well as the formation rate of EMRIs \citep[see for example,][]{Bode_2014,Naoz_2022,Naoz_2023}.} \citep{Chen_2011,Li_2015,Li_2019,Mockler_2023,Melchor_2024}.

The population of disrupted stars is likely dominated by low-mass stars, with masses $m\lesssim\Msun$, since they are more common and have a longer lifespan than more massive stars \citep{Kochanek_2016}. Yet, there are observational indications of disruptions of more massive stars, with masses up to a few solar masses \citep[e.g.,][]{Mockler_2022,Hinkle_2024,Wiseman_2024}.

We highlight the impact of mass segregation on the rates and mass distribution of the different galacto-centric transients. 
For example, if the disrupted stars of all masses were to originate from roughly the radius of influence, their relative abundances would correspond to their prevalence in the stellar population. 
However, if mass segregation occurs, massive objects occupy more tightly bound orbits.
The impact of such segregation is widely studied with regard to BHs and EMRI rates \citep[e.g.,][]{Hopman_2006b,Amaro_Seoane_2011,Aharon_2016,Raveh_2021,BroBorBon22,Rom_2024b}.
Analogously, mass segregation could increase the disruption rate of massive stars.
But, unlike BHs, the relevance of segregation for massive stars is uncertain due to their short lifetimes. 

Nonetheless, massive stars are found in the center of our galaxy, despite the local relaxation time being longer than their expected ages. This is the known ``paradox of youth'' \citep{GhezDuch_2003,LuGhez_2006,GenEisGil_2010}.
Unfortunately, the mechanisms that enable massive stars to reside deep within the sphere of influence in our own galactic center are unclear. 
Various models have been suggested; including in situ formation \citep{Levin_2003,Milosavljevic_2004}, migration within a disk or a cluster \citep{Levin_2006,Fujii_2010}, scattering by massive perturbers and binary disruptions \citep{Gould_2003,Ginsburg_2006,Perets_2007,Generozov_2020,Verberne_2025,Generozov_2025}, and a recent massive BH binary merger \citep{Akiba_2024}. 

In this work, we consider a stellar population characterized by a continuous present-day mass function (PMF), which we construct from the stellar initial mass function (IMF) and the approximate main-sequence lifetimes (Section \ref{sec:star}).
In addition to the stars, we introduce a population of stellar-mass BHs, described by a simplified power-law mass function between 10 and 30$M_\odot$ (Section \ref{sec:BH}).
Accounting for interactions between BHs and stars, we determine their steady-state distributions within the sphere of influence, assuming a relaxed cusp has formed. 
We further show that stellar-mass BHs sink into the sphere of influence, increasing their number fraction and flattening their mass function within this region (Sections \ref{sec:star_seg}--\ref{sec:BH_seg}).
Finally, in Section \ref{sec:Rates} we calculate the mass-dependent rates of TDEs, plunges, EMRIs, and QPEs, and show that transients involving massive stars and stellar-mass BHs occur more frequently than expected based on their overall abundance.
We summarize our results in Section \ref{sec:sum}.
Throughout this paper, masses are expressed in solar units.

\section{Steady-state Distributions} \label{sec:2}
We consider an SMBH, with mass $\MBH$, surrounded by a nuclear stellar cluster composed of stars and stellar-mass BHs.
We focus on the dynamics within the radius of influence, $R_h$, where the gravitational potential is dominated by the SMBH. Given the observed scaling $\MBH\propto\sigma_{h}^{4}$, where $\sigma_{h}$ is the stellar velocity dispersion \citep{Kormendy_2013}, the radius of influence is given by
\begin{equation} \label{eq:Rh}
R_h=\frac{G\MBH}{\sigma_h^2}\simeq3\ {\rm pc}\left(\frac{\MBH}{\Mmw}\right)^{1/2},
\end{equation}
where we normalize $\MBH$ to the mass of Sgr A*, $\Mmw= 4\times10^6 M_\odot$ \citep{Ghez_2008,Gillessen_2009}.

Under assumptions of spatial spherical symmetry, isotropic velocities, and weak two-body scattering dynamics, \cite{BW_76} derived a steady-state ``zero-flux'' solution for a single-mass population. 
In this solution, the phase-space distribution is given by $f(E)\propto E^p$ with $p=1/4$, which corresponds to a number density $n(r)\propto r^{-\eta}$, with $\eta=3/2+p=7/4$ (hereafter, the BW cusp).
This ``zero-flux'' solution is characterized by a vanishing particle flux and a constant energy flux \citep{Shapiro_76,Biney_Tremaine,Sari_2006,Rom_2023}.

In realistic systems, where the energy (or semimajor axis) range is finite and an absorbing boundary condition is imposed near the SMBH, the steady-state solution has a finite, nonzero particle flux. 
Yet, this flux is significantly smaller - by roughly four orders of magnitude in a Milky Way-like galaxy - than the ``naive'' estimate, given by the number of bodies over the relaxation time \citep[see][]{Rom_2023}. 
This suppression reflects the ratio of energies in the scattering region, where two-body scattering dominates over GW emission \citep{Amaro_2018,Sari_Fragione_2019,Linial_2023}.
Hence, the steady-state solution asymptotically approaches the ``zero-flux'' BW cusp.

In a following paper, \cite{BW_77} generalized their calculation for multimass groups. Assuming that the most massive objects, $m_{\max}$, are the most abundant, a ``zero-flux'' solution is satisfied when the massive group follows the single-mass BW cusp, with $p_{\max}\approx1/4$, while lighter objects, with mass $m_i<m_{\max}$, obtain shallower profiles \citep{BW_77}, satisfying
\begin{equation}\label{eq:psBW}
    \frac{p_i}{p_{\max}}\approx \frac{m_i}{m_{\max}}.
\end{equation}

However, low-mass stars are much more abundant than massive stars or compact objects; hence, they dominate the scattering and the massive objects sink toward the center of the cluster, due to dynamical friction, as encapsulated by \cite{AH_09} and \cite{Keshet_2009} mass-segregated distributions. 
\cite{Linial_2022} revisited the impact of segregation in nuclear stellar clusters and derived a steady-state ``zero-flux'' solution, taking into account the dominance of different mass groups in different energy bins. 
This framework was adapted to a two-mass model by \cite{Rom_2024b}, who used it to estimate the EMRI formation rate.

Furthermore, numerous numerical studies have examined the distributions around SMBHs, their evolution toward a steady state, and the consequent transients' formation, using N Body simulations \citep[e.g.,][]{Baumgardt_2004a,Baumgardt_2004b,Amaro_Seoane_2011,Panamarev_2019}, Monte Carlo (MC) methods \citep[e.g.,][]{Freitag_2002,Freitag_2006,Vasiliev_2015,Bar_or_2016,Balberg_2024,Zhang_2024,Zhang_2025}, or the Fokker-Planck (FP) formalism \citep[e.g.,][]{Hopman_2006b,Oleary_2009,Merritt_2015a,Aharon_2016,Vasiliev_17,BroBorBon22}. 
The formation of a BW cusp in a single-mass population surrounding an SMBH has been confirmed by multiple numerical studies since its original derivation \citep[e.g.,][]{Freitag_2002,Baumgardt_2004a,Freitag_2006,Vasiliev_2015,Merritt_2015b,Bar_or_2016}. 
In the multimass case, mass segregation of the heavier objects is generally observed, though the characteristics of the resulting distribution vary across the numerical studies, as further discussed in Section \ref{sec:BH_seg}.

\newpage
\subsection{The Present-day Stellar-mass function} \label{sec:star}
The PMF is given by \citep{Miller_1979,Kochanek_2016}
\begin{equation}\label{eq:mdist}
    \frac{\dd \Ns}{\dd m}\propto \left.\frac{\dd \Ns}{\dd m}\right|_{\rm IMF}\times\min\left\{1,\frac{\tau_{\rm nuc}}{\tau_{\rm gal}}\right\},
\end{equation}
where $\tau_{\rm gal}\sim 10\ \rm Gyr$ is the galaxy lifetime and $\tau_{\rm nuc}\sim10m^{-2.5}\ {\rm Gyr}$ is the stellar nuclear timescale over a wide range of stellar masses. 
For stars with masses $m\gtrsim40$, the stellar luminosities approach the Eddington limit \citep{Sanyal_2015} and their lifetimes become approximately constant, $\tau_{\rm nuc}(m\gtrsim40)\sim{\rm Myr}$.
Given that the total enclosed stellar mass within the sphere of influence is roughly $\MBH$ \citep{Biney_Tremaine,Merritt_2004}, we adopt a simplified normalization $\Ns(1)\approx\MBH$.

We assume a Kroupa IMF \citep{Kroupa_2001}. The broken power law of this IMF, as well as the distinction between stars with lifespans shorter and longer than the age of the galaxy, results in a broken power-law PMF, $\dd \Ns/\dd m\propto m^{-\gamma}$, with
\begin{equation}\label{eq:gamma}
\gamma\approx\left\{\def\arraystretch{1}\begin{tabular}{@{}l@{\quad}l@{}}
  $1.3$ & $0.08\leq m\leq0.5$ \\
  $2.3$ & $0.5\leq m\lesssim1$ \\
  $4.8$ & $1\lesssim m\lesssim 40$\\
  $2.3$ & $m\gtrsim 40$\\
\end{tabular}\right..
\end{equation}

Note that the short lifetimes of massive stars may prevent them from reaching a steady-state distribution, since their lifetimes can be shorter than the relaxation timescale. 
However, observations reveal a population of massive stars in the galactic center, where the local relaxation time exceeds their expected lifetimes \citep{GhezDuch_2003,LuGhez_2006,GenEisGil_2010}. 
In the absence of a widely accepted model for how these stars arrived there, we adopt a simplified approach of assuming a relaxed stellar cusp, with the mass distribution described by the PMF.
The finite lifetimes of stars are manifested in the steeper slope of the PMF for $m\gtrsim 1$ compared to the IMF (Equations \ref{eq:mdist} and \ref{eq:gamma}).

\subsection{BH mass function} \label{sec:BH}

Alongside the stars, a population of stellar-mass BHs will inevitably accumulate.
While their precise mass distribution remains uncertain, we focus on its qualitative features and its influence on the rates of the different transients.

We assume that BHs at the mass range $10\lesssim\mb\lesssim30$ follow a power-law profile 
\begin{equation}\label{eq:Nb}
    \Nb\left(\mb\right)=\fb\MBH\mb^{1-\gb}.
\end{equation}
As our fiducial values, we adopt $\gb=3$, a BH number fraction $f^{\rm IMF}_{\bullet}=10^{-3}\approx\Nb(10)/\Ns(1)$, as expected from the stellar IMF, and, consequently, $\fb=0.1$. This distribution was chosen as a simple example in which the $30\Msun$ BHs dominate the scattering. 
However, our qualitative results do not depend on the exact details of the BH mass function, as discussed in Section \ref{sec:BH_seg}.

We assume that the BH mass distribution declines significantly outside the range $10\lesssim\mb\lesssim30$ and has little influence on our results.
At the lower-mass end, we adopt a minimal BH mass of $\mb=5$ \citep[consistent with the ``lower-mass gap", see][]{LIGO_GWTC3_mass}.
This distribution is schematically presented in Figure (\ref{fig:2}).

\begin{figure}[!ht] 
    \centering
    \includegraphics[width=8.6cm,height=7.25cm]{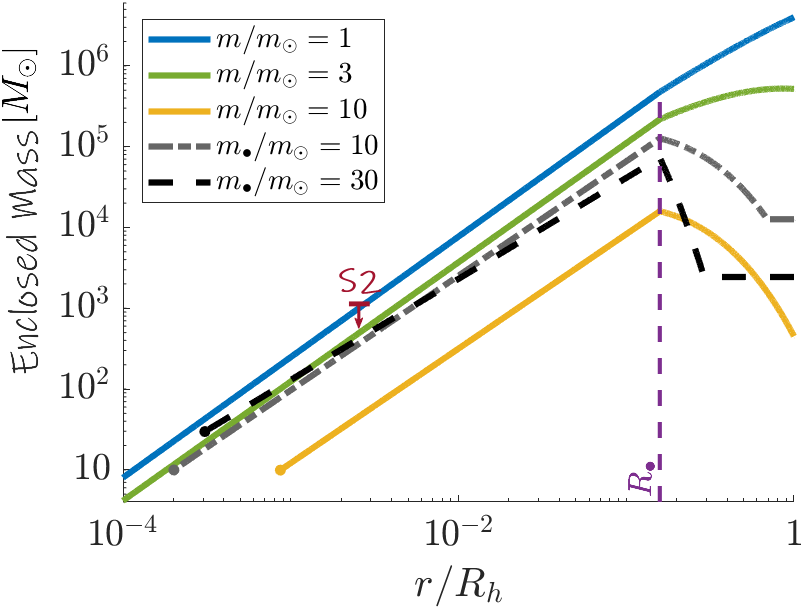}
    \caption{The enclosed mass of stars (solid colored lines) and BHs (dashed and dashed-dotted lines) as a function of semimajor axis $r$. The purple dashed vertical line marks $\Rb$ (Equation \ref{eq:Rb}), below which $30\Msun$ BHs dominate the scattering and follow the BW cusp. In this region, lighter stars and BHs follow shallower power-law profiles. At larger distances ($\Rb\lesssim r\lesssim R_h$), massive objects become exponentially rare as they segregate inward \citep[as derived by][]{Linial_2022}. The BH distribution flattens around $R_h$ due to the influx of BHs from greater distances. The red arrow indicates the observational upper limit for the extended mass enclosed within the orbit of S2 \citep{Gravity_Exmass_2024}. The distributions are truncated once a single object remains.}   
    \label{fig:1}
\end{figure}


\subsection{Stellar Segregation} \label{sec:star_seg}

Assuming that the nuclear stellar cluster is dynamically relaxed, the stellar steady-state distribution satisfies a zero particle flux \citep{BW_76,BW_77,Linial_2022}, which corresponds to a constant, nonvanishing energy flux \citep{Biney_Tremaine,Sari_Fragione_2018,Rom_2023}.

\cite{Linial_2022} derived the zero-flux, steady-state solution for a power-law mass function and observed that weighting the segregated distribution by a factor of $m^{3/2}$ reproduces the single-mass BW cusp.
Here, we offer a simple explanation for this result, which allow us to apply it for the more realistic PMF, as given by Equation (\ref{eq:mdist}).

We consider the constant energy flux through semimajor axis $r$. We assume that at each semimajor axis the energy flux is dominated by objects of mass $m$, where $m$ may be a function of $r$.
For each mass group, the energy flux induced by self-scattering (i.e., scattering between objects of the same mass) is 
\begin{equation}\label{eq:Eflux}
\begin{aligned}
    \mathcal{F}_E&\approx \frac{E(r,m)\Ns(m)}{\trlx(r,m)}
\end{aligned}    
\end{equation}
where $E(r,m)\propto m/r$ is the orbital energy, and $\trlx$ is the relaxation timescale. Furthermore, $\trlx$ is given by \citep{Biney_Tremaine,Merritt_2011}
\begin{equation} \label{eq:trlx}
\begin{aligned}
    \trlx\left(r,m\right)&\approx \frac{P(r)}{\log\Lambda}\frac{\left(\MBH/m\right)^2}{N\left(m\right)}, 
\end{aligned}
\end{equation}
where $P(r)=\sqrt{r^3/\left(G\MBH\right)}$ is the orbital period, $N(m)= m\dd N/\dd m \propto m^{1-\gamma}$, and $\log\Lambda\sim \log\left(\MBH\right)$ is the Coulomb logarithm.

Combining Equations (\ref{eq:Eflux}) and (\ref{eq:trlx}) we find the energy flux scales as $\mathcal{F}_E\propto m^3N^2r^{-5/2}$.
Maintaining a constant energy flux, with a given mass spectrum 
(e.g., Equations \ref{eq:mdist} and \ref{eq:gamma}), determines the density profile as well as the mass that dominates the scattering at each radius.
In this interpretation, the effective cross section of $m^{3/2}$, identified by \cite{Linial_2022}, arises from the combination of the scattering cross-section, $\sigma\propto m^2$, and the orbital energy, $E\propto m$.

The constant energy flux solution naturally leads to mass-segregated distribution. 
Massive stars (with $m\gtrsim1$) sink inward up to a characteristic distance $r_d(m)$ where their self-scattering dominates the energy flux. 
Using the mentioned scaling, together with the constant energy flux requirement and the stellar PMF, gives
\begin{equation}\label{eq:xmd}
    \frac{r_d(m)}{R_h}\approx m^{2-4\gamma/5},
\end{equation}
where one solar-mass stars, whose lifetime roughly equals the age of the galaxy, dominate at the radius of influence\footnote{Notably, demanding that more massive stars become the dominant scatterers at smaller distances requires $\gamma>5/2$.}. 
This result reproduces the relation obtained by the detailed calculation of \cite{Linial_2022}.
As discussed below, the presence of a stellar-mass BH population alters this behavior for stars with masses $m\gtrsim3$.

The spatial distribution of different stars and BHs is demonstrated in Figure (\ref{fig:1}), where we present the enclosed mass as a function of the semimajor axis. Evidently, low-mass stars ($m\lesssim1$) dominate the mass budget throughout most of the cluster. The contribution from more massive objects to the total enclosed mass becomes comparable to that of the low-mass stars around $r\sim10^{-4}R_h$, where only a few of these massive objects reside.

Recent observations of the Milky Way's center place an $1\sigma$ upper limit of approximately $1200\Msun$ on the extended mass enclosed within the orbit of S2 \citep[depicted by the red arrow in Figure \ref{fig:1};][]{Gravity_Exmass_2024}. 
Our segregated profile predicts an extended mass of about $\sim2000\Msun$, slightly exceeding this observational constraint. However, we expect stellar collisions to deplete the number of stars in the inner regions of the cluster, thereby lowering the extended mass within the S2 orbit. A detailed account of the impact of collisions is deferred to a subsequent paper.

\subsection{BH Segregation} \label{sec:BH_seg}
The assumption of a relaxed nuclear stellar cluster implies that massive objects can migrate into the sphere of influence during the galaxy's lifetime.
Beyond the radius of influence, the stellar distribution can be approximated as an isothermal sphere, where $N(r)\propto r$ and the relaxation timescale for objects of mass $m$ scales as $\tau_{\rm rlx}(r>R_h,m)\propto r^2/m$ \citep{Biney_Tremaine}.

Therefore, more massive objects can segregate inwards from greater distances, scaling as $r\propto m^{1/2}$.
This process increases the mass distribution of stellar-mass BHs within the sphere of influence by a factor of $\mb^{1/2}$, resulting in an effective power-law index
\begin{equation}\label{eq:tgb}
    \tgb=\gb-\frac{1}{2}.
\end{equation}

Thus, the BH number fraction inside the sphere of influence, $f^{\rm cusp}_{\bullet}$, is expected to be enhanced by $f^{\rm cusp}_{\bullet}/f^{\rm IMF}_{\bullet} \approx3$ relative to its expected value from the (nonsegregated) stellar IMF.
This enhancement impacts the expected number of EMRIs that would be observed by LISA \citep{Babak_17,Bonetti_2020,Pozzoli_2023,Rom_2024b}.
Additionally, this incoming flux of BHs flattens their spatial distribution near the radius of influence, as schematically illustrated in Figure (\ref{fig:1}).

Although massive stars may segregate from beyond the sphere of influence as well, we do not account for this effect in their distribution due to their short lifetimes and the uncertainty surrounding their migration mechanisms.

Within the sphere of influence, the BHs sink, due to dynamical friction from the stars, to a characteristic distance $\Rb$, where BHs with masses $\tmb\approx30$ become the dominant scatterers.
This distance is determined by comparing the energy flux of BHs (analogous to Equation \ref{eq:Eflux}) to the constant energy flux set by one solar-mass stars at the radius of influence. This yields
\begin{equation}\label{eq:Rb}
\frac{\Rb}{R_h}=\left(\frac{\Nb(\tmb)^2\tmb^3}{M^2}\right)^{2/5}=\fb^{4/5}=0.16.
\end{equation}

As presented in Figure (\ref{fig:1}), from $\Rb$ inward, the $30\Msun$ BHs follow the single-mass BW cusp, while lighter stars and BHs settle into shallower profiles (according to Equation \ref{eq:psBW}).
Thus, massive stars with $m\gtrsim3$ will accumulate around $\Rb$, where they will be scattered by the BHs, rather than efficiently segregating to $r_d(m)\lesssim\Rb$.
The critical stellar mass is given by $\widetilde{m}=\fb^{2/\left(5-2\gamma\right)}\approx3$, satisfying $r_d(\widetilde{m})\approx\Rb$. 
Therefore, we conclude that in the segregated distribution, stars of mass $m$ are concentrated around a characteristic distance, $\widetilde{r}_d(m)$, given by
\begin{equation} \label{eq:rdm_t}
    \widetilde{r}_d\approx\left\{\def\arraystretch{1}\begin{tabular}{@{}c@{\quad}l@{}}
  $R_h$ & $m\lesssim1$ \\
  $r_d(m)$ & $1\lesssim m\lesssim3$ \\
  $\Rb$ & $m\gtrsim3$ \\
  \end{tabular}\right.,
  \end{equation}
where $r_d(m)$ and $\Rb$ are given in Equations (\ref{eq:xmd}) and (\ref{eq:Rb}), respectively.

Our results are weakly sensitive to the shape of the poorly constrained BH mass function. 
In our fiducial model, we assume that the heavy $\approx30\Msun$ BHs are sufficiently abundant to dominate the scattering, namely,
$\Nb(30)/\Nb(10)\gtrsim3^{-3/2}$.
The observations of the LIGO-Virgo-KAGRA (LVK) collaboration \citep{LIGO_GWTC3_mass} suggest that this is indeed the case, if the binary merger population is a good representative of the overall mass distribution of stellar-mass BHs.
Specifically, we adopt a power-law BH mass function with $\gb=3$ (which leads to an effective slope $\tgb=5/2$ due to sinking of BHs from outside the sphere of influence, see Equations \ref{eq:Nb} and \ref{eq:tgb}). This is the the steepest power-law profile where $30\Msun$ BHs dominate the scattering. Therefore, a shallower profile would result in limited changes to our results. 

However, if the more massive stellar-mass BHs are less abundant, e.g., a steeper mass function with $\gb>3$, 
then the $10\Msun$ BHs become the dominant scatterers at $\Rb$, assuming the total BH number fraction remains fixed.
In this case, the scattering timescale at $r\lesssim\Rb$ increases by a factor of $3^{1/2}$, compared to our fiducial model (see Equation \ref{eq:trlx2}). The impact of heavier stellar-mass BHs then becomes subdominant, as they concentrate at smaller radii, $r < \Rb$, where fewer objects reside and their contribution to the overall dynamics is diminished.

\subsection{Comparison with numerical studies} \label{sec:BH_seg_comp}
Our result for the stellar and BH distributions, derived from the condition of constant energy flux (as presented in Figure \ref{fig:1}), indicates that stellar-mass BHs sink and become the dominant scatterers within $\Rb\sim0.1R_h$ (Equation \ref{eq:Rb}). In this inner region ($r\lesssim\Rb$), the BHs form a BW cusp, with a power-law slope $\eta_\bullet=-(d\log n/d\log r)=1.75$, while lighter objects, with mass $m_\ell$, develop shallower profiles, $\eta_\ell=1.5+m_\ell/(4m_\bullet)\approx1.5$, as given by Equation (\ref{eq:psBW}). 

Dynamically, we expect somewhat flatter profiles than noted above, especially in the inner regions of the cluster, if there is a sink term, such as stellar collisions or loss-cone depletion \citep{Cohn_1978,Bar_or_2016}, which is not included in our model.
Observationally, such flattened profiles, with slopes $\eta=1.1-1.4$, have been reported for the stellar population in our own Galactic center \citep{Cano_2018,Schodel_2018}.

Several numerical studies have found that BHs develop a BW cusp, while lighter objects exhibit shallower profiles, in the inner regions of the nuclear stellar cluster ($r\lesssim 0.1R_h$), in agreement with our prediction. 
Table \ref{tab:comp} summarizes the power-law slopes inferred by the various numerical studies discussed below.

\cite{Freitag_2006} used a spherical stellar dynamical MC code that includes weak and strong scattering, stellar evolution, collisions, and an approximate treatment of loss-cone depletion. Their results show the expected cusp slopes for both the BHs and the stars (see their Figs. 8 and 10 for two- and multimass models, respectively).
\cite{BroBorBon22} obtained similar slopes using a FP approach for a two-mass system of $1\Msun$ stars and $10\Msun$ BHs (which we refer to as the fiducial two-mass system), including scattering and loss-cone depletion (see their Figure 4).
\cite{Panamarev_2019} conducted direct N body simulations with one million particles, accounting for stellar evolution and binary stars, and found consistent slopes for BHs, WDs, and stars (see their Figure 3). \cite{Baumgardt_2018} also employed a direct N body simulation (with $5\times10^4$ particles), including scattering, stellar evolution, and star formation. While their results show that BHs follow the steepest profile, the inferred slope ($\eta_\bullet=1.55$) is shallower than values reported in other studies. 
This discrepancy may be due to the fact that the reported slope reflects an average over both BHs and lighter neutron stars (NSs).

\cite{Balberg_2024} further confirmed these trends through MC simulations of the fiducial two-mass system, incorporating weak scattering, GW emission, loss-cone depletion, and stellar collisions (see their Figs. 3 and 4).
Their results support two additional predictions of our model \cite[and that of][for the two-mass case]{Rom_2024b}. First, stars near $R_h$ follow a BW cusp, whereas BHs develop a steeper profile, although their reported slope of $\eta_\bullet \approx 2$ is shallower than our analytical prediction (see discussion below). 
Second, they confirm that the radial extent of the BH cusp depends on the BH number fraction $f_\bullet$, in excellent agreement with our analytical estimate \citep[see Table \ref{tab:comp}, Equation (\ref{eq:Rb}) here and Equation (10) in][]{Rom_2024b}.

Other numerical studies find steeper cusps for the BHs, $\eta_\bullet\approx2$.
For instance, \cite{Hopman_2006b} numerically solved the FP equation for a multimass population composed of stars, WDs, NSs, and BHs (see their Figure 1). 
\cite{Oleary_2009} extend this approach to models with power-law BH mass function (see their Figure 1). They found that BHs dominate within $0.1R_h$, where the lighter components follow shallower profiles consistent with Equation (\ref{eq:psBW}), though with $p_{\max}\approx0.5$ rather than $p_{\max}\approx0.25$ \citep[as in][]{BW_77}.

\cite{Preto_Amaro_Seoane_2010} carried out N body simulation (with $\sim10^5$ particles) and FP calculation for the fiducial two-mass system, approximately reproducing either the \cite{BW_77} solution or the \cite{AH_09} solution depending on the BH number fraction (see their Figure 2).
\cite{Aharon_2016} applied the FP method to models with different star formation histories, including stars, WDs, NSs, $10\Msun$ and/or $30\Msun$ BHs. They found that BHs consistently develop the steepest profile, with the exact slope depending on the assumed star formation scenario (see their Figs. 1 and 2)\footnote{An exception is a model including in situ star formation beyond the sphere of influence, in which all species, including BHs, exhibit flatter slopes of $\eta\approx1$.}.
Finally, \cite{Zhang_2025} explored a suite of two-mass models using an MC method, incorporating two-body scattering, the evolving stellar potential, and SMBH growth. Their results are consistent with the general trend of steep BH cusps and shallower profiles for lighter species (see their Figure 9).

The discrepancies in the reported slopes, as presented in Table \ref{tab:comp}, may arise from limitations in numerical resolution or small number statistics, as the innermost regions of the cluster are sparsely populated.
Additionally, in the outer region of the cluster, $\Rb\lesssim r\lesssim R_h$, our model predicts a very steep profile for the massive objects. In the multimass case, they become exponentially rare \citep{Linial_2022}, while in a two-mass model their distribution follows a power law with $\eta_\bullet=4$ for $10M_\odot$ \citep{Rom_2024b}, and even steeper slopes for more massive BHs, following Equation \ref{eq:psBW}.
These predictions are steeper than those of \cite{AH_09} in their ``strong segregation'' regime, as we argue for a constant energy flux while they assume a constant particle flux. 

To our knowledge, such steep stellar-mass BH profiles near the radius of influence have not been observed in numerical simulations. This may be due to the fact that these extreme slopes occur only within a narrow radial range (as illustrated in Figure \ref{fig:1}), making them challenging to resolve numerically. Nevertheless, the presence of steep slopes in the cluster outskirts could bias the inferred average slope if the BH distribution is fitted with a single power law over the entire sphere of influence.

\begin{deluxetable*}{lccc} \label{tab:comp}
\tabletypesize{\footnotesize}
\tablewidth{0.6\columnwidth}
\tablecaption{Comparison of BH and stellar distributions in nuclear stellar clusters from numerical studies \label{tab:glossary}}
\tablehead{
\colhead{Reference} & \colhead{$\eta_\bullet$\tablenotemark{\footnotesize a}} & \colhead{$\eta_\ell$\tablenotemark{\footnotesize b}} & \colhead{Radial Extent\tablenotemark{\footnotesize c}}
}
\startdata
{\bf This work} & {\bf 1.75} & {\bf $\approx$ 1.5} & {\bf $r\lesssim0.16R_h$ }\\ \hline
\cite{Freitag_2006} & $1.7-1.8$ & $1.3-1.4$ & $r\lesssim0.15R_h$\\ \hline
\cite{Hopman_2006b} & $\approx2$ & $1.4-1.5$  & $r\lesssim R_h$\\ \hline
\cite{Oleary_2009} & $\approx2$ & $\approx1.5$  & $r\lesssim 0.1R_h$\\ \hline
\cite{Preto_Amaro_Seoane_2010} & $2.1\ (1.8)$\tablenotemark{\footnotesize d} & $1.5$  & $r\lesssim R_h\ (0.1R_h)$\tablenotemark{\footnotesize d}\\ \hline
\cite{Aharon_2016} & $1.9-2.3$ & $1.3-1.5$ & $r\lesssim 10^{-2}R_h$ \\ \hline
\cite{Baumgardt_2018} & $1.55$ & $1-1.2$ & $r\lesssim R_h$\\ \hline
\cite{Panamarev_2019} & $1.72\pm0.04$ & $0.9-1.2$ & $r\lesssim0.5R_h$ \\ \hline
\cite{BroBorBon22} & $1.7$ & $1.3$ & $r\sim 10^{-3}R_h$ \\ \hline
\cite{Balberg_2024} & $1.75$ & $1.475$ & $r\lesssim0.05R_h\ (0.2R_h)$\tablenotemark{\footnotesize e}\\ \hline
\cite{Zhang_2025} & $1.75-2.2$ & $1.1-1.4$ & $0.01R_h\lesssim r \lesssim R_h$ \\ \hline
\enddata
\tablenotetext{a}{Slope of the distribution, $\eta_\bullet=-(d\log n_\bullet/d\log r)$, for the dominant stellar-mass BHs.}
\tablenotetext{b}{Slope of the distributions of the lighter populations, which may include stars, WDs, and NSs, depending on the specific model.}
\tablenotetext{c}{Radial extent over which the reported power-law slopes are estimated.}
\tablenotetext{d}{Value depends on the BH number fraction, $f_\bullet\approx2.5\times10^{-3}$ ($f_\bullet\approx0.5$).}
\tablenotetext{e}{Value depends on the BH number fraction, $f_\bullet\approx10^{-3}$ ($f_\bullet\approx10^{-2}$).}
\end{deluxetable*}

\section{Transients mass-dependent rates} \label{sec:Rates}
Based on the steady-state distributions of stars and BHs, 
the transient formation rates can be estimated using the two-body scattering induced flux \citep{Wang_2004,Hopman_2005,Stone_2016}. 
Here, we generalize this concept by taking into account the mass segregation as a function of radius, and arrive at
\begin{equation}\label{eq:rate0}
    \frac{\dd\Gamma}{\dd m}\approx\frac{\dd N/\dd m}{\trlx(r,m)\log\Lambda'},
\end{equation}
The logarithmic term in the denominator originates from the diffusive flux in angular momentum\footnote{Our rate estimation assumes an empty loss-cone dynamics. For further details see \cite{Lightman_1977,Vasiliev_2013,Alexander_2017}.} and is given by the log of the ratio between the orbital semi-major axis to the loss-cone size, $\left(R_t/r\right)$ for stars or $\left(4R_s/r\right)$ for BHs. 
This introduces a weak dependence on the energy (and, in the case of stars, on the stellar tidal radius), which we simplify by approximating the logarithmic term as a constant, equal to the Coulomb logarithm (i.e., $\log\Lambda'\approx\log\Lambda$). 

We estimate the rate of each transient type and object mass by evaluating Equation (\ref{eq:rate0}) at the radius where it is maximal, as described below.
Additionally, we define the enhancement factor $\mathcal{Q}$ as the ratio between the rate of a specific transient, involving stars or BHs, and their abundance in the total population
\begin{equation}\label{eq:Q}
    \mathcal{Q}=P(R_h)\MBH\frac{\dd\Gamma/\dd m}{\dd N/\dd m}.
\end{equation}
The normalization factor, $P(R_h)\Mmw$, is roughly the number of one solar-mass stars at the radius of influence divided by their tidal disruption rate. It is set such that $\mathcal{Q}_{TDE}\left(1\right)=1$.

\subsection{TDE rate} \label{sec:TDErate}

The majority of disrupted stars with masses in the range $1\lesssim m\lesssim3$ originate from the vicinity of $r_d(m)$, as defined by Equation (\ref{eq:xmd}), where they dominate the scatterings. In this region, the relaxation timescale (Equation \ref{eq:trlx}) is
\begin{equation} \label{eq:trlx2}
    \trlx\left(m\right)\sim 10 m^{-\gamma/5}\left(\frac{\MBH}{\Mmw}\right)^{5/4} {\rm Gyr}.
\end{equation}
At smaller distances, both their number and the two-body scattering flux decrease as a power law, and at larger distances their number is exponentially suppressed \citep{Linial_2022}.

Therefore, using Equations (\ref{eq:mdist}), (\ref{eq:rate0}), and (\ref{eq:trlx2}), the mass-dependent tidal disruption rate is given by
\begin{equation}\label{eq:rate}
    \begin{aligned} 
    \Rate{TDE}&\approx\frac{\dd\Ns/\dd m}{\trlx\left(m\right)\log\Lambda}\approx\frac{m^{-4\gamma/5}}{P(R_h)}\\
    &\approx 3\times10^{-5}m^{-4\gamma/5}\left(\frac{\MBH}{\Mmw}\right)^{-1/4}{\rm yr^{-1}}.
    \end{aligned}
\end{equation}
Consequentially, the enhancement factor (defined in Equation \ref{eq:Q}) is 
\begin{equation}\label{eq:Q2}
    \mathcal{Q}_{TDE}=m^{\gamma/5}.
\end{equation}
Given that $\gamma\approx5$ for this stellar-mass range (Equation \ref{eq:gamma}), the TDE rate scales roughly as $\dd\Gamma/\dd m\propto m^{-4}$ and the enhancement factor is $\mathcal{Q}_{TDE}\simeq m$.
Notably, the TDE rate, as well as the rates of EMRIs and plunges (calculated below), scales with the SMBH mass as $\Gamma\propto \MBH^{-1/4}$ \citep[in agreement with previous results, e.g.,][]{Hopman_2005,Kochanek_2016,Yao_2023}.

As discussed in Section \ref{sec:BH_seg}, the presence of a BH population modifies the cusp within $R_{\bullet}$.
The transition of massive stars ($m\gtrsim\widetilde{m}$) from being primarily self-scattered to being scattered by BHs results in a sharp increase in the enhancement factor\footnote{This increase originates from the shorter relaxation timescale when scattered by the more massive BHs compared to the stars, $\trlx(\widetilde{m})/\trlx(\tmb)\approx3$, as evident from Equations (\ref{eq:trlx2}) and (\ref{eq:trlx3}).} around $\Rb$, 
$\delta\mathcal{Q}\approx\left(\tmb/\widetilde{m}\right)^{1/2}\approx3$. 
Therefore
\begin{equation}\label{eq:QTDE10}
    \begin{aligned}
    \mathcal{Q}_{TDE}(m\gtrsim\widetilde{m})&=\mathcal{Q}_{TDE}(\widetilde{m})\delta\mathcal{Q}\\
    &=\tmb^{1/2}\fb^{1/\left(5-2\gamma\right)}\approx9.
    \end{aligned}
\end{equation}

At the other end of the mass spectrum, the segregation of low-mass stars, with $m\lesssim1$, is negligible. As a result, disrupted stars of such masses typically originate near the radius of influence, where they are scattered by the one solar-mass stars. Their disruption rate follows their mass function, $\left.\dd\Gamma/\dd m\right|_{TDE}\propto m^{-\gamma}$ and thus $\mathcal{Q}_{TDE}\left(m\lesssim1\right)=1$. 

\begin{figure*}[ht!]
    \centering
    \includegraphics[width=\textwidth,height=0.6\textwidth]{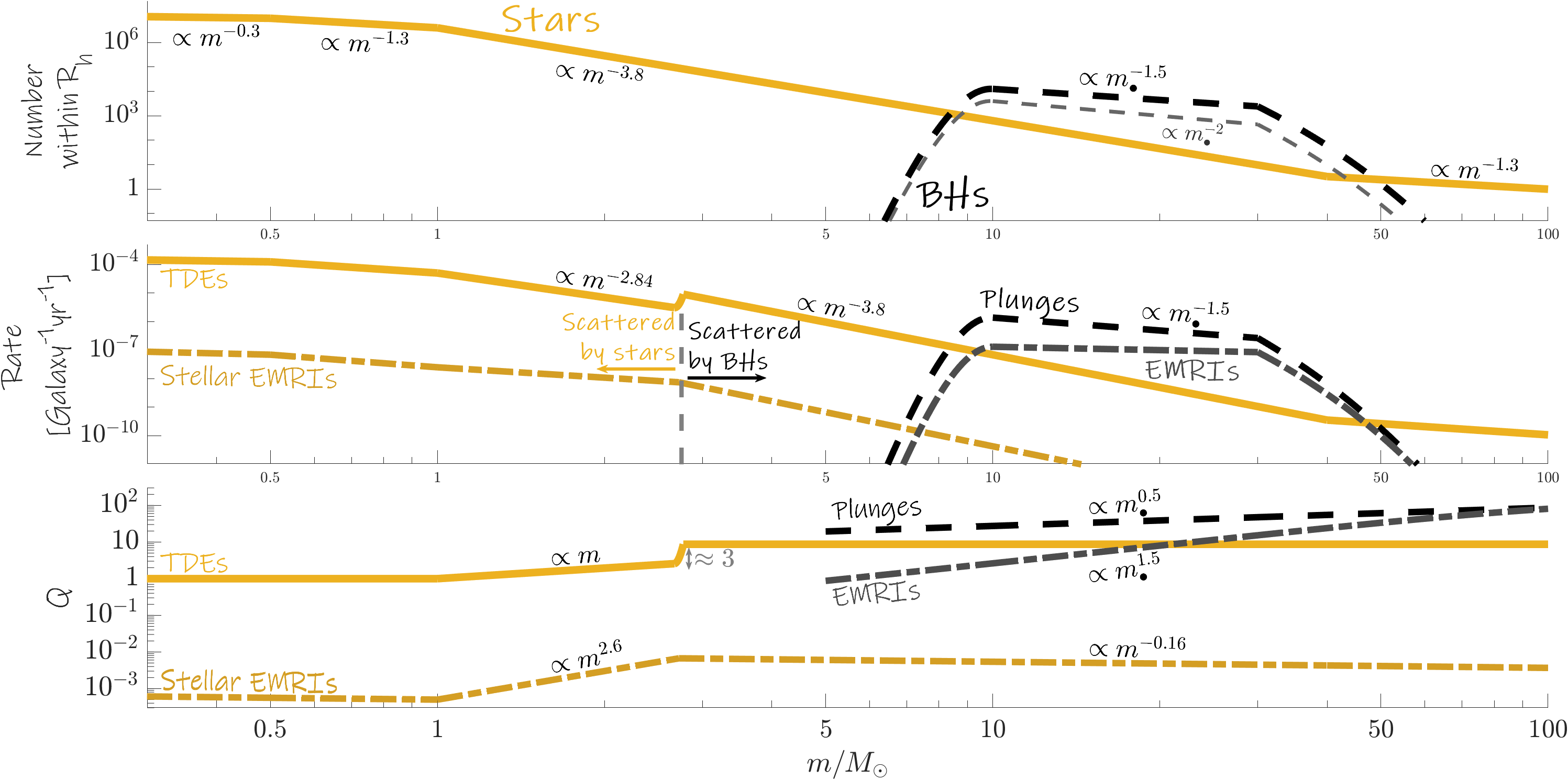}
    \caption{Distributions for stars (yellow lines) and BHs (dashed black lines). 
    {\bf Top panel}: The number, $N(m)=m\dd N/\dd m$, of stars and BHs within the sphere of influence. The stars follow the stellar PMF (Equation \ref{eq:mdist}). We show our assumed BH mass distribution (thin gray dashed line, Equation \ref{eq:Nb}), and its modification due to segregation from beyond the sphere of influence (thick black dashed line, Equation \ref{eq:tgb}). 
    {\bf Second panel}: The rates of TDEs, plunges, EMRIs, and stellar EMRIs (sEMRIs). The dashed vertical gray line separates masses that are primarily scattered by stars ($m\lesssim3$) from those mostly scattered by $30\Msun$ BHs. 
    {\bf Bottom panel}: The enhancement factor $\mathcal{Q}$ for the different transients. 
    Mass segregation within the sphere of influence generally leads to over-representation of transients involving massive stars, resulting in $\mathcal{Q}\approx 9$ for TDEs with $m\gtrsim3$ (Equation \ref{eq:QTDE10}). The sEMRIs are suppressed compared to TDEs, but also show over-representation of massive stars. However, for very massive stars ($m\gtrsim3$), $\mathcal{Q}$ weakly decreases (Equation \ref{eq:QsE}), reflecting the variation of the tidal radius with stellar mass.
    For TDEs, $\mathcal{Q}$ sharply increases around $m\approx3$, as indicated by the gray arrow, since these stars become mostly scattered by BHs.
    For BHs, segregation beyond the sphere of influence takes place as well, leading to an enhancement factor that increases with mass. For EMRIs, $\mathcal{Q}$ rises more steeply due to the additional dependence of the EMRI critical radius on the BH mass (Equation \ref{eq:Rc}).}
    \label{fig:2}
\end{figure*}

The enhancement factor of stellar disruptions is illustrated in Figure (\ref{fig:2}) and can be summarized as follows:
\begin{equation}\label{eq:Q_tot}
    \mathcal{Q}_{TDE}\approx\left\{\def\arraystretch{1}\begin{tabular}{@{}c@{\quad}l@{}}
  $1$ & $m\lesssim1$ \\
  $m$ & $1\lesssim m\lesssim3$\\
  9 & $m\gtrsim3$\\
\end{tabular}\right..
\end{equation}

\subsection{Plunge rate} \label{sec:Plungerate}

The stellar-mass BHs are predominantly scattered by the $\tmb\approx30$ BHs near $\Rb$. 
Therefore, the relevant scattering timescale (Equation \ref{eq:trlx}) for the plunge rate is given by
\begin{equation} \label{eq:trlx3}
\begin{aligned}
    \tau_{{\rm rlx},\bullet}\left(r\right)&\approx \frac{P\big(r\big)}{\log\Lambda}\frac{\left(\MBH/\tmb\right)^2}{\Nb\left(\tmb\right)\left(r/\Rb\right)^{5/4}} \\
    &= \frac{P\big(R_h\big)\MBH}{\log\Lambda}\tmb^{-1/2}\left(\frac{r}{R_h}\right)^{1/4}\\
    &\sim 1\ {\rm Gyr}\left(\frac{r}{\Rb}\right)^{1/4}\left(\frac{\MBH}{\Mmw}\right)^{5/4},
\end{aligned}
\end{equation}
where we used Equations (\ref{eq:trlx}), (\ref{eq:Nb}), and (\ref{eq:Rb}).

The resulting plunge rate, for BHs in the mass range $10\lesssim\mb\lesssim30$, is
\begin{equation}\label{eq:rate2}
    \begin{aligned}
        \Rate{Plunge}&\approx \frac{\dd\Nb/\dd\mb}{\log\Lambda\tau_{{\rm rlx},\bullet}(\Rb)}\approx\frac{\fb^{4/5}\tmb^{1/2}\mb^{-\tgb}}{P\left(R_h\right)}\\
        &\approx7\times10^{-8}\left(\frac{\mb}{10}\right)^{-5/2}\left(\frac{\MBH}{\Mmw}\right)^{-1/4}{\rm yr^{-1}}.
    \end{aligned}
\end{equation}

Consequently, the enhancement factor is
\begin{equation}\label{eq:Qplunge}
    \mathcal{Q}_{Plunge}\approx27\left(\frac{\mb}{10}\right)^{1/2}.
\end{equation} 
This scaling reflects the flattening of the BH mass spectrum caused by the segregation into the sphere of influence (see Section \ref{sec:BH_seg}). 
It applies to both lighter and heavier BHs\footnote{For massive stellar-mass BHs, with $\mb\gtrsim125$, The critical radius $R_{c}$ (Equation \ref{eq:Rc}), which differentiate between plunges and EMRIs, exceeds $\Rb$ (Equation \ref{eq:Rb}), the radius around which the BHs concentrate. Therefore, such BHs are more likely to form EMRIs than undergo a plunge.\label{fn:msbh}}, since they are mostly scattered by BHs with mass $\tmb$ around $\Rb$, leading to $\left.\dd\Gamma/\dd m\right|_{Plunge}\propto\mb^{-\tgb}$.

\subsection{EMRI rate} \label{sec:EMRIrate}
In order to estimate the EMRI rate, we first determine the characteristic radius, $R_{c}$, from which they originate\footnote{Notably, this dichotomy between EMRI and plunge progenitors becomes less pronounced for lower-mass SMBHs, with $\MBH\lesssim10^5 M_\odot$ \citep[see][]{Qunbar_2023,Mancieri_2024}} \citep{Hopman_2005,Amaro_2018,Sari_Fragione_2019}.
This radius is defined such that the GW timescale and the scattering timescale are comparable for orbits with $r=R_{c}$ and $r_p=4R_s$ \citep[for further details see][]{Rom_2024b}. 
It is given by
\begin{equation}\label{eq:Rc}
\begin{aligned}
    \frac{R_{c}}{R_h}&\approx\left(\frac{\mb}{16\sqrt{2}\log\Lambda\tmb^{1/2}}\right)^{4/5}
    \approx0.02\left(\frac{\mb}{10}\right)^{4/5}.
\end{aligned}
\end{equation}
This value is slightly lower than previous estimates \citep[e.g.,][]{Hopman_2005,Rom_2024b,KaurPerets_2024} due to the BH mass distribution we consider. Specifically, in our model, stellar-mass BHs are scattered by more massive $30\Msun$ BHs, rather than the commonly assumed $10\Msun$ BHs. The presence of more massive scatterers reduces the scattering timescale, leading to the smaller critical distance. 

The EMRI rate for BHs in the mass range $10\lesssim\mb\lesssim30$ is therefore
\begin{equation}\label{eq:rate3}
    \begin{aligned}
        \Rate{EMRI} &\approx \frac{\dd\Nb/\dd\mb\left(R_{c}/\Rb\right)^{3/2-\mb/(4\tmb)}}{\log\Lambda \tau_{{\rm rlx},\bullet}\left(R_{c}\right)}\\
        &\approx 2\times10^{-7}\mb^{1-\tgb}\left(\frac{125}{\mb}\right)^{\mb/150}\\
        &\quad\times \left(\frac{\MBH}{\Mmw}\right)^{-1/4}{\rm yr^{-1}},
    \end{aligned}
\end{equation}
and the corresponding enhancement factor is
\begin{equation}\label{eq:QEMRI}
\begin{aligned}
    \mathcal{Q}_{EMRI}&=0.07\mb^{3/2}\left(\frac{125}{\mb}\right)^{\mb/150}\approx2.6\left(\frac{\mb}{10}\right)^{3/2}.
\end{aligned}
\end{equation}

Thus, we get that for $10\Msun$ BHs the EMRI rate is $\approx10^{-7}\ {\rm yr^{-1}}$ and they are enhanced by a factor of $\mathcal{Q}_{EMRI}(10)\approx2.7$.
The scaling $\mathcal{Q}_{EMRI}\propto\mb^{3/2}$ stems from two contributions: a factor of $\mb^{1/2}$ due to the segregation beyond the sphere of influence (similar to the case of plunges), and a factor of $\mb$ from the dependence of the critical radius on the BH mass (Equation \ref{eq:Rc}).

Furthermore, using Equations (\ref{eq:rate2}) and (\ref{eq:rate3}), we determine the EMRI-to-plunge ratio, a key parameter for predicting the number of sources detectable by LISA \citep[see][]{Babak_17,Rom_2024b}
\begin{equation}\label{eq:E2P}
\begin{aligned}
    \frac{\left.\dd\Gamma/\dd m\right|_{EMRI}}{\left.\dd\Gamma/\dd m\right|_{Plunge}}&=\left(\frac{R_{c}}{\Rb}\right)^{\left(5-\mb/\tmb\right)/4}\\
    &\approx\left(\frac{\mb}{125}\right)^{1-\mb/150}.\\
\end{aligned}
\end{equation}

For $10\Msun$ BHs, Equation (\ref{eq:E2P}) predicts an EMRI-to-plunge ratio of $\approx0.1$. 
This estimate is consistent with the fiducial value considered by \cite{Babak_17} and is about $1.5$ times lower than the result of \cite{Rom_2024b}, which was based on a two-mass model of the nuclear stellar cluster.
The lower ratio arises because, in this work, the $10\Msun$ BHs are scattered by heavier $30\Msun$ BHs.

\subsection{Stellar EMRI rate} \label{sec:sEMRIrate}
Finally, we consider the case where a main-sequence star slowly descends toward the SMBH due to GW emission, i.e., an EMRI of a main-sequence star rather than of a BH. 
In this case, the critical semimajor axis is found by equating the minimal periapsis to the tidal radius \citep{Linial_2023}, in contrast to $4R_s$ as in the BH case. This yields
\begin{equation}\label{eq:Rc_star}
\begin{aligned}
    \frac{R_{\star,c}}{R_h}&\approx \left(190\tmb^{1/2}\log\Lambda\right)^{-4/5}m^{-0.13}\left(\frac{\MBH}{\Mmw}\right)^{4/3}\\
    &\approx6\times10^{-4}m^{-0.13}\left(\frac{\MBH}{\Mmw}\right)^{4/3}.
\end{aligned}
\end{equation}

The resulting main-sequence stellar EMRI (sEMRI) rate is
\begin{equation}\label{eq:rate4}
    \begin{aligned}
        \Rate{sEMRI}&\approx \frac{\dd \Ns/\dd m\left(R_{\star,c}/\widetilde{r}_d\right)^{3/2}}{\log\Lambda\tau_{{\rm rlx},\bullet}\left(R_{\star,c}\right)}\\
        &\approx 10^{-8}m^{-\gamma-0.16}\left(\frac{\MBH}{\Mmw}\right)^{17/12}{\rm yr^{-1}}\\
        &\ \ \times\left\{\def\arraystretch{1}\begin{tabular}{@{}c@{\quad}l@{}}
  $1$ & $m\lesssim1$ \\
  $m^{6\gamma/5-3}$ & $1\lesssim m\lesssim3$ \\
  $16$ & $m\gtrsim3$ \\
  \end{tabular}\right.,
    \end{aligned}
\end{equation}
where we assume that stars of mass $m$ are concentrated around $\widetilde{r}_d(m)$, as given in Equation (\ref{eq:rdm_t}).
The calculation is further simplified by assuming that for $r\lesssim\widetilde{r}_d$ the stellar number density scales as $r^{-3/2}$, neglecting the mass-dependent correction to the power of the density profile (Equation \ref{eq:psBW}).

The weak $m^{-0.16}$ scaling of the sEMRI rate arises from the critical radius $R_{\star,c}$ (Equation \ref{eq:Rc_star}). This reflects the tidal radii variation across different stellar masses, and, hence, implicitly depends on the stellar-mass-radius relation. 
The additional mass-dependent term in the intermediate mass range, $1\lesssim m\lesssim3$, arises from the segregated location of these stars.

The sEMRI rate in our model is smaller compared to calculations assuming a single-mass nuclear stellar cluster \citep{Linial_2023} or a two-mass model \citep{Kaur_2024}. 
This is because, in our model, the stars are scattered by more massive BHs, which shortens the scattering timescale and reduce the critical distance for EMRI formation, shifting it to regions where stars are less abundant.
 
From Equations (\ref{eq:Q}) and (\ref{eq:rate4}), the sEMRI enhancement factor is
\begin{equation}\label{eq:QsE}
\begin{aligned}
    \mathcal{Q}_{sEMRI}\approx&5\times10^{-4}m^{-0.16}\times\left\{\def\arraystretch{1}\begin{tabular}{@{}c@{\quad}l@{}}
  $1$ & $m\lesssim1$ \\
  $m^{2.8}$ & $1\lesssim m\lesssim3$ \\
  $16$ & $m\gtrsim3$ \\
  \end{tabular}\right..
\end{aligned}
\end{equation}

Besides being mHz GW sources, relevant for space-based GW observatories, sEMRIs have recently gained attention as a potential origin of the observed QPEs \citep{Zhao_2021,Linial_2023,LuQua_2023,Linial_Metzger_2023,Nicholl_2024}.
Specifically, \cite{Linial_Metzger_2023} associated QPEs with the interaction between a sEMRI and a TDE accretion disk. 

Given this interpretation, the QPE rate can be estimated by combining the TDE rate (Equation \ref{eq:rate2}), the sEMRI rate (Equation \ref{eq:rate4}), and the characteristic EMRI lifetime, $\tau_{GW}\sim m^{-1}{\rm Myr}$ \citep{Linial_Metzger_2023,Kaur_2024}. 
This yields a characteristic rate $\sim10^{-6} \rm yr^{-1}$ for one solar-mass stars, roughly compatible with the observational inferred rate \citep{Arcodia_2024}. 

The QPE rate scales as $m^{-1}\left.\dd\Gamma/\dd m\right|_{sEMRI}$, suggesting that low-mass stars (with $m\lesssim1$) may be preferable candidates, based only on rate considerations.

We note that beyond the dynamical processes considered in this work, the effects of stellar binaries and collisions should be incorporated for a full analysis of the sEMRI formation mechanism. For example, binary tidal break-up may be a more efficient formation channel \citep{Linial_2023,LuQua_2023}, while collisions are expected to reduce the abundance of stars on tightly bound orbits, suppressing the sEMRI rate \citep{Sari_Fragione_2019,Rose_2023,Balberg_2023,Balberg_2024}.

\section{Discussion \& Summary} \label{sec:sum}

We study the distributions of stars and BHs in nuclear stellar clusters surrounding SMBHs. 
We analytically study the impact of two-body scattering and the resulting mass segregation, and self-consistently derive the steady-state distributions for a population of stars and stellar-mass BHs. Based on these distributions, we estimate the formation rate of various transients and show that mass segregation generally enhances the fraction of transients involving massive stars and stellar-mass BHs relative to their overall abundances.

We assume that the cluster is dynamically relaxed, with stars following a PMF that accounts for their finite lifetimes, resulting in a steeper slope for $m\gtrsim 1$ relative to the stellar IMF. 
This working assumption is motivated by observations of massive stars in the central parsec of our galaxy \citep{GhezDuch_2003,LuGhez_2006,GenEisGil_2010}. However, it should be revisited once the presence of massive stars deep in the sphere of influence, the so-called ``paradox of youth'', is better understood.

Additionally, we assume that $30\Msun$ BHs are sufficiently abundant to dominate over all other stellar-mass BHs. 
This is inspired by the emerging BH mass spectrum, inferred from the observations of the LVK collaboration \citep{LIGO_GWTC3_mass}.
Given this assumption, the specific details of the BH distribution - taken here as a power-law toy model with tails extending to both low and high mass ends - slightly affect the quantitative rate estimates but do not alter the qualitative trends discussed in this work.
Furthermore, in the case that the $10\Msun$ BHs are the dominant scatterers, with the overall BH number fraction held fixed, the scattering timescale and the characteristic distances for EMRI and sEMRI formation would change only by factors of order-unity. 
Notably, while the formation rates depend on the shape of the BH mass function, as they scale linearly with the number of BHs at each mass, this dependence is factored out in the enhancement factor $Q$. The enhancement trends, pointed out in our analysis,  reflect dynamical effects and are largely insensitive to the specific form of the mass function.

We present a simple derivation of the segregated steady-state distributions for both the stars and the stellar-mass BHs, accounting for their mutual interactions. 
These distributions are obtained using our generalized formulation of the constant energy flux condition, which enables the derivation of several new results.

We show that the number of BHs within the sphere of influence increases by a factor of $\approx 3$ relative to the expectations based on the stellar IMF, since their relaxation time is shorter than the age of the galaxy, even beyond the sphere of influence, enabling them to sink inward from larger radii.
Furthermore, we find that BHs dominate the scattering within $\Rb\sim0.1R_h$. Therefore, in this inner region, they form a BW cusp. This prediction is consistent with the findings of several numerical studies \citep[e.g.,][]{Freitag_2006,Panamarev_2019,BroBorBon22,Balberg_2024}.

Taking into account the formation of transients, we estimate the EMRI-to-plunge ratio and show that scattering by more massive BHs reduces the EMRI rate.
This is because the scattering by heavier BHs shifts the EMRIs critical radius toward smaller distances, where the BH population is less abundant.
Nonetheless, the segregated BH distribution leads to an over-representation of EMRIs and plunges involving massive stellar-mass BHs, with enhancement factor scaling as  $\mathcal{Q}_{EMRI}\propto \mb^{3/2}$ and $\mathcal{Q}_{Plunge}\propto \mb^{1/2}$, respectively.
These results are of particular interest because they influence the expected number of EMRIs to be measured by LISA \citep{Babak_17,Amaro_2023,Rom_2024b}. 

Regarding TDEs, we show that massive stars, with masses $1\lesssim m\lesssim3$, have an enhanced disruption rate, compared to their abundance in the stellar population, by roughly a factor of $\mathcal{Q}_{TDE}=m$.
This enhancement arises from segregation within the sphere of influence, leading to an accumulation of massive stars at smaller distances, where the scattering timescale is shorter. 
The disruption of more massive stars, with $m\gtrsim3$, is enhanced by a constant factor of $\approx9$, as they are predominantly scattered by BHs around a characteristic distance of $\Rb\sim0.1R_h$.

Observationally, the tidal disruption rate of massive stars is not well constrained.
Mass estimates of disrupted stars, based on the light curves of approximately 20 observed TDEs, generally indicate a preference for lower-mass stars, with a probability tail extending to more massive ones \citep{Mockler_2019, Ryu_2020m, Zhou_2021}. 
Additionally, \cite{Hinkle_2024} and \cite{Wiseman_2024} detected several long-lived luminous flares, which they suggest may originate from the disruption of stars with masses of $m\approx3-10$.
\cite{Mockler_2022} identified a preference for the disruption of moderately massive stars, with $m\gtrsim2$, relative to their prevalence in the stellar population, based on the nitrogen-to-carbon abundances in a few observed TDEs.
Our analysis shows that this trend is expected due to mass segregation, which enhances the rate of such events.

Finally, we estimate the sEMRI rate, showing that it favors low-mass stars ($m\lesssim1$). We find that the sEMRI rate is reduced compared to its expected rate from single-mass models \citep[e.g.,][]{Linial_2023}, yet remains marginally consistent with the observationally inferred QPE rate \citep{Arcodia_2024}.
However, a more detailed analysis, including the effects of stellar binaries and collisions \citep{Rose_2023,Balberg_2023,Balberg_2024}, is necessary for accurately determining the sEMRI rate.

Future observations, combined with a more detailed model of stellar and BH distributions, will clarify the extent of relaxation and mass segregation in galactic centers. These efforts will also determine whether the observed disruptions of massive stars can be fully attributed to mass segregation, or if additional factors, such as a recent star formation burst, are required.

\begin{acknowledgments}
The authors would like to thank Itai Linial, Brenna Mockler, Nicholas Stone, Tsvi Piran, Vikram Ravi, and Jean Somalwar for useful discussions. This research was partially supported by an ISF grant, an NSF/BSF grant, an MOS grant, and a GIF grant. B.R. acknowledges support from the Milner Foundation.
\end{acknowledgments}

\bibliography{main}{}
\bibliographystyle{aasjournal}

\end{document}